\documentclass[twocolumn,tighten]{aastex63} %

\usepackage[version=4]{mhchem}
\usepackage{amsmath}

\received{29 Jan 2020}
\revised{}
\accepted{}
\submitjournal{ApJ Letters}

\shorttitle{Interstellar comet 2I/Borisov seen by MUSE/TRAPPIST}
\shortauthors{Bannister, Opitom, Fitzsimmons et al.}

\graphicspath{{./}{figures/}}

\begin{document}

\title{Interstellar comet 2I/Borisov as seen by MUSE: C$_2$, NH$_2$ and red CN detections}

\correspondingauthor{Michele Bannister}
\email{michele.t.bannister@gmail.com}

\author[0000-0003-3257-4490]{Michele T. Bannister}
\affiliation{Astrophysics Research Centre, School of Mathematics and Physics, Queen's University Belfast, Belfast BT7 1NN, United Kingdom}
\affiliation{School of Physical and Chemical Sciences -- Te Kura Mat\={u}, University of Canterbury, Private Bag 4800, Christchurch 8140, New Zealand}

\author[0000-0002-9298-7484]{Cyrielle Opitom} 
\affiliation{ESO (European Southern Observatory) - Alonso de Cordova 3107, Vitacura, Santiago Chile}
\affiliation{Institute for Astronomy, University of Edinburgh, Royal Observatory, Edinburgh EH9 3HJ, UK}

\author[0000-0003-0250-9911]{Alan Fitzsimmons}
\affiliation{Astrophysics Research Centre, School of Mathematics and Physics, Queen's University Belfast, Belfast BT7 1NN, United Kingdom}

\author[0000-0001-9784-6886]{Youssef Moulane}
\affiliation{ESO (European Southern Observatory) - Alonso de Cordova 3107, Vitacura, Santiago Chile}
\affiliation{STAR Institute, Universit\'e de Li\`ege, All\'ee du 6 aout, 19C, 4000 Li\`ege, Belgium}
\affiliation{Oukaimeden Observatory, Cadi Ayyad University, Marrakech, Morocco}

\author[0000-0001-8923-488X]{Emmanuel Jehin}
\affiliation{STAR Institute, Universit\'e de Li\`ege, All\'ee du 6 aout, 19C, 4000 Li\`ege, Belgium}

\author{Darryl Seligman}
\affiliation{Department of Astronomy, Yale University, 52 Hillhouse Ave., New Haven, CT 06517}

\author{Philippe Rousselot} 
\affiliation{Institut UTINAM UMR 6213, CNRS, Univ. Bourgogne Franche-Comté, OSU THETA, BP 1615, 25010 Besançon Cedex, France}

\author[0000-0003-2781-6897]{Matthew M.~Knight}
\affiliation{Department of Physics, United States Naval Academy, 572C Holloway Rd, Annapolis, MD 21402, USA}
\affiliation{University of Maryland, Department of Astronomy, College Park, MD 20742, USA}

\author[0000-0001-8617-2425]{Michael Marsset}
\affiliation{Department of Earth, Atmospheric and Planetary Sciences, MIT, 77 Massachusetts Avenue, Cambridge, MA 02139, USA}

\author[0000-0003-4365-1455]{Megan E. Schwamb}
\affiliation{Astrophysics Research Centre, School of Mathematics and Physics, Queen's University Belfast, Belfast BT7 1NN, United Kingdom}

\author[0000-0003-2354-0766]{Aur{\'e}lie Guilbert-Lepoutre} 
\affiliation{Laboratoire de G{\'e}ologie de Lyon, LGL-TPE, UMR 5276 CNRS / Universit{\'e} de Lyon / Universit{\'e} Claude Bernard Lyon 1 / ENS Lyon, 69622 Villeurbanne, France}

\author{Laurent Jorda} 
\affiliation{Aix Marseille Univ, CNRS, LAM, Laboratoire d'Astrophysique de Marseille, Marseille, France}

\author{Pierre Vernazza} 
\affiliation{Aix Marseille Univ, CNRS, LAM, Laboratoire d'Astrophysique de Marseille, Marseille, France}

\author{Zouhair Benkhaldoun}
\affiliation{Oukaimeden Observatory, Cadi Ayyad University, Marrakech, Morocco}

\begin{abstract}

We report the clear detection of C$_2$ and of abundant NH$_2$ in the first prominently active interstellar comet, 2I/Borisov.
We observed 2I on three nights in November 2019 at  optical wavelengths 4800--9300 \AA with the Multi-Unit Spectroscopic Explorer (MUSE) integral-field spectrograph on the ESO/Very Large Telescope. 
These data, together with observations close in time from both 0.6-m TRAPPIST telescopes, provide constraints on the production rates of species of gas in 2I's coma. 
From the MUSE detection on all epochs of several bands of the optical emission of the C$_2$ Swan system, a rich emission spectrum of NH$_2$ with many highly visible bands, and the red (1-0) bandhead of CN, together with violet CN detections by TRAPPIST, we infer production rates of
$Q$(C$_2$) = $1.1\times10^{24}$ mol s$^{-1}$,
$Q$(NH$_2$) = $4.8\times10^{24}$ mol s$^{-1}$ 
and $Q$(CN) = $(1.8\pm0.2)\times 10^{24}$ mol s$^{-1}$.
In late November at 2.03~au, 2I had a production ratio of C$_2$/CN$=0.61$, only barely carbon-chain depleted, in contrast to earlier reports measured further from the Sun of strong carbon-chain depletion.
Thus, 2I has shown evolution in its C$_2$ production rate: a parent molecule reservoir has started sublimating.
At $Q$(NH$_2$)/$Q$(CN) = 2.7, this second interstellar object is enriched in NH$_2$, relative to the known Solar System sample.

\end{abstract}

\keywords{}

\section{Introduction} 
\label{sec:intro}

Interstellar objects offer an unprecedented opportunity to closely study remnants of the star and planet formation process from systems other than our own. The comae of interstellar comets permit detailed characterization of the chemical composition of their volatile material and the physical properties of their dust. Although planet formation is a robust process, the architectures of extrasolar planetary systems are remarkably diverse; many are substantially different to the architecture of our own system. Characterizing planetesimals from other stars will contextualize the formation and evolution of our own system, through comparison to the Solar System minor planet populations.

2I/Borisov\footnote{Initially designated C/2019 Q4 (Borisov); MPEC 2019-R106 \url{https://www.minorplanetcenter.net/mpec/K19/K19RA6.html}} is the second interstellar object ever detected entering our Solar System, and the first to exhibit distinct cometary activity. In contrast to 2I, the first interstellar object, 1I/`Oumuamua, displayed a marked lack of a coma in deep stacks of images \citep[e.g.][]{Meech2017}. The sublimation of volatiles from 1I was only inferred from the minor non-gravitational acceleration in its outbound trajectory \citep{Micheli:2018}. However, 1I had distinctive characteristics even from the upper-limit measurements of its volatiles: for instance, it needed to be depleted in \ce{CN} by at least a factor of 15 relative to typical abundances in comets, if the non-gravitational forces were due to the sublimation of \ce{H2O} \citep{ISSITeam:2019}. Only two previous Solar System comets have been seen with comparable \ce{CN} depletions, and those were hypothesized to be potentially captured interstellar objects \citep{Fink:1992, Schleicher:2008}.
2I has a definitively interstellar origin, with an escape velocity, $v_{\infty} = 32.304 \pm 0.001$ km/s, on a hyperbolic trajectory with an eccentricity of $3.3575 \pm	0.0003$\footnote{JPL Horizons, 20 Nov 2019: \url{https://ssd.jpl.nasa.gov/sbdb.cgi?sstr=2I;old=0;orb=0;cov=0;log=0;cad=0\#elem} }. 
This is well within the expected distribution of values for the eccentricity of a detectable active interstellar object, given the inherent bias in the Sun's gravitational focusing of the trajectories of these objects \citep{Engelhardt:2017}. 2I entered the Solar System at an inclination of $44^{\circ}$ to the ecliptic, and its inbound trajectory passed  outside of the orbit of Mars at 2.006 au, near the peak of detectable perihelion values for active objects \citep{Engelhardt:2017}. 
Hubble Space Telescope optical imaging indicates a nuclear size of 0.2--0.5 km for 2I \citep{Jewitt:2019b}, supported by upper limits from ground-based infrared imaging \citep{Bolin:2019,Lee:2019}. 
From this, the size distribution of interstellar objects is such that the discovery of 2I is consistent with extrapolations based on 1I \citep{Jewitt:2019b}.

The cometary activity of 2I provides the first opportunity for an extensive compositional study of a volatile-rich interstellar object. A dusty coma with an initial colour of $g'-r' = 0.63 \pm 0.03$ and D-type visible spectrum implied a first-look similarity to long-period comets \citep{Guzik:2019, de-Leon:2019, Jewitt:2019, Bolin:2019, Yang:2019}.
High-airmass spectra soon after the 2019 August 30 discovery showed the presence of \ce{CN} \citep{Fitzsimmons:2019}. 
Ongoing upper limits on \ce{C2} \citep{Opitom:2019-borisov, Kareta:2019}, edging towards weak detections \citep{Lin:2019}, have been at levels implying that 2I is carbon-depleted, similar to $\sim$30\% of the Solar System comets \citep{Ahearn:1995,Cochran:2012}.
The brightening rate in \citet{Ye:2019}'s pre-discovery observations of 2I back to December 2018 appears flat, as does the more densely sampled brightening rate since discovery \citep[e.g. Fig. 3,][]{Opitom:2019-borisov}. 
This is consistent with the behaviour of dynamically new comets, and much less steep than that of Jupiter-family comets \citep{Whipple:1978}.
2I's lack of detection in \citet{Ye:2019}'s earlier imaging implies that activity driven by species more volatile than \ce{H2O}, such as \ce{CO} or \ce{CO2}, began between $\sim 7.8-8.5 {\, \rm au}$. 
\ce{H2O} sublimation has apparently driven 2I's activity from around 6~au inward \citep{Fitzsimmons:2019, Jewitt:2019, Ye:2019}, with detections of water products increasing toward perihelion. 
No evidence of water-ice rich grains was found on October 9 at 2.41 au by \cite{Yang:2019}.
By October 11, the \ce{[OI]} 6300 \AA\ detection of \citet{McKay:2019} implied, if it is produced solely by water dissociation, an \ce{H2O} production rate for 2I of ($6.3\pm1.5)\times10^{26}$ mol s$^{-1}$.
A marginal 18-cm OH production rate was measured from October 2-25 with the Nan\c{c}ay radio telescope at $(3.3 \pm 0.9) \times 10^{27}$ mol s$^{-1}$ \citep{Crovisier:2019}.
{\it Swift}/UVOT measurements show an increasing water production rate, going from (7.0 $\pm$ 1.5) $\times 10^{26}$ to (10.7 $\pm$ 1.2) $\times 10^{26}$ mol s$^{-1}$ from November 1 to December 1, 2019 \citep{Xing:2020}.

We report the first observations from our ongoing detailed and comprehensive characterization of 2I's composition and activity over time, using the Multi-Unit Spectroscopic Explorer \citep[MUSE;][]{Bacon:2010} on the 8.2~m UT4 of the ESO/Very Large Telescope (VLT).
Constraining the composition and nature of 2I's volatiles and the production of its dust particles across the spatial distribution of the coma requires both spectroscopy and imaging \citep[e.g.][]{Opitom:2019-C2016ER61}.
MUSE's wide-field integral-field spectroscopy, with medium spectral resolution and sensitivity to faint and diffuse targets, is ideally suited to the task. 
Its 4800--9300 \AA\ wavelength range contains several molecular and atomic species seen in emission in normal Solar System comets \citep[e.g.,][]{Feldman:2004}.
The species that directly leave the nucleus of the comet are termed ``parent" species, and are  non-detectable in the optical.
They fragment to detectable ``daughter" species, by many and varied paths, predominantly photodissociation.
The common daughter species \ce{C2} and \ce{CN} have strong emission bands, with \ce{C2} (0-0) at bandhead 5165 \AA\ and the red \ce{CN} system A$^2\Pi$-X$^2\Sigma^+$, with bandheads at 9109 \AA\ and 7822 \AA. Several bands of the amino radical \ce{NH2}, the primary dissociation product of ammonia (\ce{NH3}), are also covered, such as the (0,3,0) band at 6300--6340 \AA. 
Although forbidden oxygen is encompassed in the wavelength range (e.g. at 6363 \AA), MUSE's resolution is not sufficient to isolate it from its overlap with \ce{NH2}, and thus we do not place any constraint on \ce{H2O} production.
Our future reports will provide simultaneous maps of the distribution in the coma of multiple gas species and dust, which we have sampled frequently throughout 2I's perihelion passage. 

In these November 2019 high-airmass and near-twilight observations, the slowly brightening comet is too faint for high-quality gas maps.
Here we report detections, initial production rates and relative abundances of three gas species: \ce{C2}, \ce{NH2}, and \ce{CN}.
These species are commonly seen in the decades of measurements of Solar System comets \citep[e.g.][]{Ahearn:1995, Fink:2009, Cochran:2012, Hyland:2019}. Our results mark the first detection of \ce{NH2} in 2I, and a definitive detection of \ce{C2} in respect to previous searches \citep{Fitzsimmons:2019, Opitom:2019-borisov, Kareta:2019, Lin:2019}.
Observations by both of the TRAnsiting Planets and PlanetesImals Small Telescope \citep[TRAPPIST;][]{Jehin:2011}, made simultaneously or near in time, provide supporting constraints on the violet CN and an upper limit on the \ce{C2} production rates.

\section{Observations and data reduction}
\label{sec:obs}

\begin{deluxetable*}{cccccccc}
\tabletypesize{\footnotesize}
\tablecolumns{7}
\tablecaption{Observations of 2I/Borisov by MUSE/VLT and TRAPPIST \label{tab:observations} }
\tablehead{
\colhead{Time} \vspace{-0.2cm} &Instrument/ & \colhead{Time} & \colhead{Exposure} & Airmass & \colhead{$r_{H}$} & \colhead{$\Delta$} & \colhead{$\alpha$} \\ 
\colhead{UT} & Filter &\colhead{MJD} & \colhead{(s)} & & \colhead{(au)} & \colhead{(au)} & \colhead{(degrees)} 
}
\startdata
2019-Nov-10 05:05 & TN/CN &2458797.70833 &1500 & 1.41 & 2.102 & 2.285 & 25.64 \\
2019-Nov-14 08:03 & MUSE &2458801.83542 & 600 & 2.07 & 2.078 & 2.230 & 26.26 \\
2019-Nov-14 08:19 & MUSE &2458801.84653 & 600 & 1.86 & 2.078 & 2.230 & 26.26 \\
2019-Nov-15 08:02 & MUSE &2458802.83472 & 600 & 2.04 & 2.072 & 2.217 & 26.40 \\
2019-Nov-15 08:18 & MUSE &2458802.84583 & 600 & 1.84 & 2.072 & 2.217 & 26.40 \\
2019-Nov-17 05:00 & TN/CN &2458804.70833 &1500 & 1.45 & 2.061 & 2.191 & 26.65 \\
2019-Nov-25 05:15 & TN/C$_2$ &2458812.71885 &1500 & 1.45 & 2.029 & 2.101 & 27.58 \\
2019-Nov-25 05:40 & TN/CN &2458812.72980 &1500 & 1.38 & 2.029 & 2.101 & 27.58 \\
2019-Nov-26 07:17 & MUSE &2458813.80347 & 600 & 2.12 & 2.025 & 2.091 & 27.69 \\
2019-Nov-26 07:34 & MUSE &2458813.81528 & 600 & 1.89 & 2.025 & 2.091 & 27.69 \\
2019-Nov-26 07:45 & TS/C$_2$ &2458813.82291 &1500 & 1.60 & 2.025 & 2.091 & 27.68 \\
2019-Nov-26 07:46 & MUSE &2458813.82361 & 600 & 1.76 & 2.025 & 2.091 & 27.69 \\
2019-Nov-26 08:03 & MUSE &2458813.83542 & 600 & 1.61 & 2.025 & 2.091 & 27.69 \\
2019-Nov-26 08:10 & TS/CN &2458813.84375 &1500 & 1.44 & 2.024 & 2.090 & 27.69 \\
\enddata
\tablecomments{$r_H$ and $\Delta$ are the heliocentric and geocentric distances, $\alpha$ the phase angle.\\ 
The MUSE November 14 observations were off-centred in the 1 arcmin$^2$ field of view, though contained all the apparent coma.
}
\end{deluxetable*}

\subsection{MUSE}

We observed 2I/Borisov with the MUSE integral-field spectrograph on VLT's UT4 at Paranal, Chile, on 2019 November 14, 15, and 26, as detailed in Table~\ref{tab:observations}. These measurements are the first part of our ongoing campaign to regularly monitor 2I through its perihelion passage\footnote{ESO Director's Discretionary program 2103.C-5070.}, providing a record of the comet's composition and ejected dust, across a range of viewing angles and distances from the Sun. The coma of 2I has thus far been comparatively compact in our observations; the entire coma fits well within the $1\times1$ arcmin$^2$ MUSE WFM-mode field of view. 
MUSE's FOV is sampled at 0.2{\arcsec} spatial resolution and operates with a spectral resolution of $R \simeq 3000$ at very high throughput \citep{Bacon:2010}.

Our observing strategy at each epoch is based on previous successful observations of comets with MUSE \citep[e.g. C/2015 ER61;][]{Opitom:2019-C2016ER61}. 
We acquire four 600 s exposures, with VLT UT4 tracked at the comet's on-sky rate of motion. We apply small dithering and a rotation of 90$^\circ$ between exposures. This allows later reduction of the signature of the detector gaps, streaked background stars, and better IFU-to-IFU uniformity in the combined data cubes. 
Sky exposures of 180 s were acquired, in an exposure sequence of OSOOSO (O = on source, S = sky), for additional clean sky emission measurements.
A spectrophotometric standard star was observed on the same night for later flux calibrations; LDS749B at airmass $X=1.2$ on Nov 14, and CPD-69 177 at $X=1.5$ on Nov 15, and at $X=1.4$ on Nov 26.
All observations were made in clear or photometric conditions, with 0.7-0.9{\arcsec} seeing.
The November 14-15 observations had a high sky background from the nearly full Moon $80-90^\circ$ distant.

Dark subtraction, flat-fielding, telluric correction, and cube reconstruction of the MUSE observations were made with its data reduction pipeline\footnote{\url{http://www.eso.org/sci/software/pipelines/muse/muse-pipe-recipes.html}} \citep{Weilbacher:2016}. 
Using the in-frame sky for simultaneous sky measurement was more accurate than sky subtraction from the sky exposures. The sky was thus estimated directly from zones covering about 20\% of the science exposures. For more details, we refer the reader to the MUSE pipeline manual. For the data from 14 and 15 November, we report only the first two exposures of each night. The final two exposures during twilight on 14 and 15 November had sufficiently strong sky background to be inadequately corrected by the pipeline, and we do not consider them further in this first analysis.
The data cubes were flux calibrated and the telluric correction was made using the spectrophotometric standard star observations from that night. 
Because of the differential tracking and the lack of point-like sources in the field of view, the cubes obtained over the same night could not easily be re-combined using the MUSE pipeline. We thus reduced the cubes separately and combined the extracted spectra. 

\subsection{TRAPPIST}

Near simultaneous observations were made with both 0.6-m TRAPPIST-North (TN) and TRAPPIST-South (TS) robotic telescopes located in the Atlas Mountains in Morocco and ESO/La Silla Observatory in Chile \citep{Jehin:2011}. TS and TN are equipped with 2k$\times$2k CCD cameras and HB comet narrow-band filters \citep{Farnham2000} with a field of view of $\sim$ 20$\arcmin \times$20$\arcmin$. The images were binned 2$\times$2, resulting in a plate scale of 1.2$\arcsec$/pixel. The narrow-band TS data for November 14--15 had low signal-to-noise, as the comet was at high airmass with full moon contamination, but good measurements were acquired with TN on November 10 and 17 with the CN filter. Data were acquired on November 25 with TN, and simultaneously with MUSE on November 26 with TS, in both cases under clear conditions and using both C$_2$ and CN filters (Table~\ref{tab:observations}). 
The data reduction of the TRAPPIST narrow band images followed standard procedures using frequently updated master bias, flat, and dark frames. The dust contamination was removed, the sky contamination subtracted, and the flux calibration performed using regularly updated zero points based on observations of photometric standard stars, as described in \citet{opitom2016} and \citet{Moulane2018}.

\section{Analysis and results}
\label{sec:results}

\begin{figure*}
    \includegraphics[width=\textwidth]{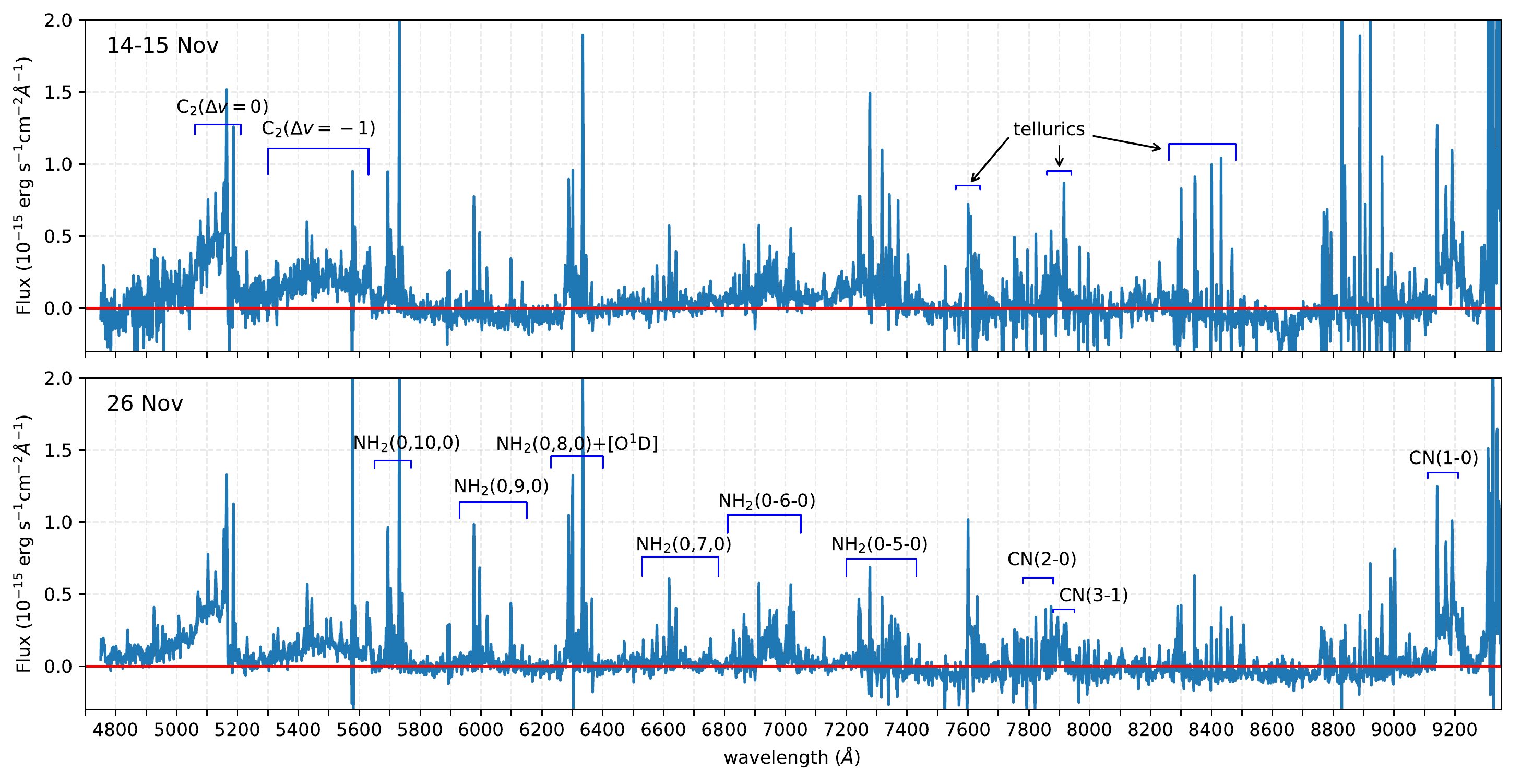}
    \caption{Spectra of the interstellar comet 2I/Borisov from two epochs: 14-15 November (combined due to bright moon conditions) and 26 November 2019, 
    observed by the MUSE VLT integral-field spectrograph. On both co-added epochs, the full spectra are extracted over a 10{\arcsec} radius aperture centred on the comet. 
    Light reflected from 2I's dust has been subtracted using a spectral-slope-corrected continuum of 67P \citep{Guilbert-Lepoutre:2016}. Several features are visible in both epochs, including the emission lines of the Swan system of \ce{C2}, many lines of \ce{NH2} through the bands 10, 9, 8 and 7, and red CN. Gas species and some telluric emission lines, or terrestrial atmospheric lines, are indicated above the spectra. The majority of the non-labeled lines in the red region of the spectrum may be attributed to tellurics.}
    \label{fig:spectrum}
\end{figure*}

\begin{deluxetable}{lccc}
\tabletypesize{\footnotesize}
\tablecolumns{4}
\tablecaption{Molecular fluorescence scattering efficiencies $g$ and Haser model scale-lengths for parent and daughter species $l_p$, $l_d$ at $r_H=1$ au, used with $v=0.5$ km/s}
\label{tab:model_parms}
\tablehead{
\colhead{Band}
& \colhead{$g$}
& \colhead{$l_p$}
& \colhead{$l_d$} \\
\colhead{}
& \colhead{ergs s$^{-1}$mol$^{-1}$} 
& \colhead{km}  
& \colhead{km}
}
\startdata
C$_2(\Delta v=0)$ & $4.5\times10^{-13}$ & $2.2\times10^{4}$ & $6.6\times10^4$ \\ 
NH$_2$(0-4-0) & $9.2\times10^{-15}$ & $4.1\times10^{3}$ & $6.2\times10^{4}$ \\ 
CN(1-0) & $9.1\times10^{-14}$ & $1.3\times10^{4}$ & $2.1\times10^{5}$ \\ 
CN(0-0) & $2.6\times10^{-13}$ & $1.3\times10^{4}$ & $2.1\times10^{5}$ \\
\enddata
\tablecomments{All $g$-factors are assumed to scale as $r_{H}^{-2}$.  
All scalelengths are assumed to scale as $r_{H}^{2}$, and are from \citet{Ahearn:1995} for \ce{C2} and CN. 
}
\end{deluxetable}

\begin{deluxetable*}{cc|cc|cc|ccc}
\tabletypesize{\footnotesize}
\tablecolumns{9}
\tablecaption{Flux from gas species detected in the coma of 2I/Borisov and their production rates in 2019 November
\label{tab:rates}
}
\tablehead{
\colhead{Date} 
& \colhead{Instrument}  
& \colhead{\ce{C2(\Delta v=0)} } 
& \colhead{Q(\ce{C2})} 
& \colhead{\ce{NH2}(0-10-0)} 
& \colhead{Q(\ce{NH2})} 
& \colhead{CN(1-0)}  
& \colhead{CN(0-0)} 
& \colhead{Q(CN)} \\
\colhead{} 
& \colhead{}
& \colhead{} 
& \colhead{$\times10^{24}$} 
& \colhead{$\times10^{-15}$} 
& \colhead{$\times10^{24}$} 
& \colhead{$\times10^{-14}$}
& \colhead{$\times10^{-16}$} 
& \colhead{$\times10^{24}$} \\  
\colhead{} 
& \colhead{}
& \colhead{ergs s$^{-1}$cm$^{-2}$} 
& \colhead{mol s$^{-1}$} 
& \colhead{ergs s$^{-1}$cm$^{-2}$} 
& \colhead{mol s$^{-1}$} 
& \colhead{ergs s$^{-1}$cm$^{-2}$}
& \colhead{ergs s$^{-1}$cm$^{-2}$} 
& \colhead{mol s$^{-1}$}  
}
\startdata
Nov. 10 & TN & \nodata & \nodata & \nodata & \nodata & \nodata & 6.6 &$1.9\pm0.5$ \\
Nov. 14-15 & MUSE & $6.8\times 10^{-14}$ & $1.1 $ & $1.4 \pm 0.1$  & $4.2 $ & $3.2\pm 0.3$ & \nodata  & $1.8\pm0.2$ \\
Nov. 17 & TN & \nodata & \nodata & \nodata & \nodata & \nodata & 7.2 & $1.9\pm0.5$  \\
Nov. 25 & TN &$<4.4\times 10^{-16}$ &  $<2.0$ & \nodata & \nodata & \nodata & 6.2 & $1.6\pm0.5$ \\
Nov. 26 & MUSE & $7.3\times10^{-14}$ & $1.1 $  & $1.5\pm0.1$ & $4.8 $ & $3.4\pm 0.3$ & \nodata & $1.8\pm0.2$\\
Nov. 26 & TS &$<4.3\times 10^{-16}$ &$<2.0$ & \nodata & \nodata & \nodata & 6.9 & $1.5\pm0.5$ \\
\enddata
\tablecomments{The measured \ce{C2} and CN fluxes differ between TRAPPIST and MUSE because of the measurement technique. TN/TS fluxes are per arcsec$^2$ measured at a projected radial distance from the nucleus equivalent to 10,000 km at the comet. MUSE fluxes are total integrated fluxes through a circular aperture of 10{\arcsec} radius.}
\end{deluxetable*}

In Fig.~\ref{fig:spectrum}, we present two MUSE spectra of the interstellar comet 2I/Borisov, each extracted over a 10{\arcsec} radius circular aperture. 
The spectra from 14 and 15 November are extracted from the two darkest-time cubes on each epoch, and subsequently co-added.
Those from 26 November are from each of the four exposure cubes acquired that night, and then co-added.
Cometary spectra are a superposition of the many emission bands from molecular gas, and the continuum of sunlight reflected from the dust in the coma. 
MUSE observations of 67P (where only dust was detected: \citet{Guilbert-Lepoutre:2016}) were corrected for the difference in spectral slope between the two comets, and used to subtract the dust-reflected sunlight contribution, producing the spectra shown in Fig.~\ref{fig:spectrum}.
All of the most common Solar System comet gas species with emission bands in the MUSE wavelength coverage are present. The spectra exhibits the distinctive shape of \ce{C2}($\delta \nu=0)$ emission with its bandhead at 5165 \AA, many notably bright red lines of \ce{NH2}, including those at 5733, 5976, 5995, 6021, 6334 {\AA} \citep{Cochran:2002}, the cometary [OI] forbidden lines overlapping with \ce{NH2} at 6363\AA, and the red CN $(1-0)$ band around 9140 \AA.  Residuals from the sky subtraction in the red part of the spectrum are particularly apparent in the Nov 14-15 data, where the telluric \ce{O2} band has been over-corrected. 

The dust continuum in the MUSE spectra was 
fit using spectral regions free of significant gas-emission between 4750--9000 \AA. 
We omit the measurement on November 14; the dust continuum shows large non-linear variations as a function of wavelength, possibly due to problems with sky subtraction due to the high airmass of observation, and we treat this measurement as unreliable. 
The spectral slope of the reddened continuum on November 26 was measured as $11.0\pm 0.2$ \%/$10^3$ \AA. 
This is similar to the spectral slope measured at red wavelengths by \cite{de-Leon:2019} on September 13, and so the dust colour of 2I appears consistent.

In Table~\ref{tab:rates}, we present the flux and gas production rates in the coma of 2I for several nights spanning our observations in November 2019, and
we consider each observed gas species in turn below. 
For the MUSE spectra, the gas species' integrated flux was retrieved for each continuum-subtracted emission band (cf. Fig.~\ref{fig:spectrum}).
We then used the parameters given in Table~\ref{tab:model_parms} to convert the fluxes into column densities, as per the standard equation \citep[e.g. Eq. 1,][]{Beaver:1990}:
\begin{equation}
    N = \frac{4\pi F}{\Omega g(r_{H})}
      = 6.81\times10^{11}\frac{Fr_{H}^{2}}{g(1\mathrm{au})\theta^2}
\end{equation}
where $N$ is the column density, $F$ the emission band's flux, $\Omega$ the solid angle of the telescope's aperture, $r_H$ the heliocentric distance, $g$ the fluorescence scattering efficiency, and $\theta$ the diameter of the measurement aperture (here 10{\arcsec}).
Using these column densities, we computed 2I's gas production rates (Table~\ref{tab:rates}) using a Haser model \citep{Haser:1957}. 
Note that while a few bands of \ce{CO+} are located in the MUSE spectral range, there is too much structure in the \ce{C2} to reliably detect \ce{CO+} at present. Also, \ce{C3} and \ce{N2+} emission are both too far to the blue to see in MUSE data.
At present, given the medium spectral resolution of MUSE, we do not identify any unusual/unidentified lines.
For the TRAPPIST narrow-band images, the gas production rates were derived from median radial brightness profiles for the CN and \ce{C2} images. The R broadband continuum filter images were used to remove the dust contamination. The fluxes were then converted to column densities and the profiles were fitted with a Haser model at 10,000 km from the nucleus (corresponding to 6.9 arcsec), using the same parameters as for the MUSE spectra (Table~\ref{tab:model_parms}).

\begin{figure}
    \includegraphics[width=\columnwidth]{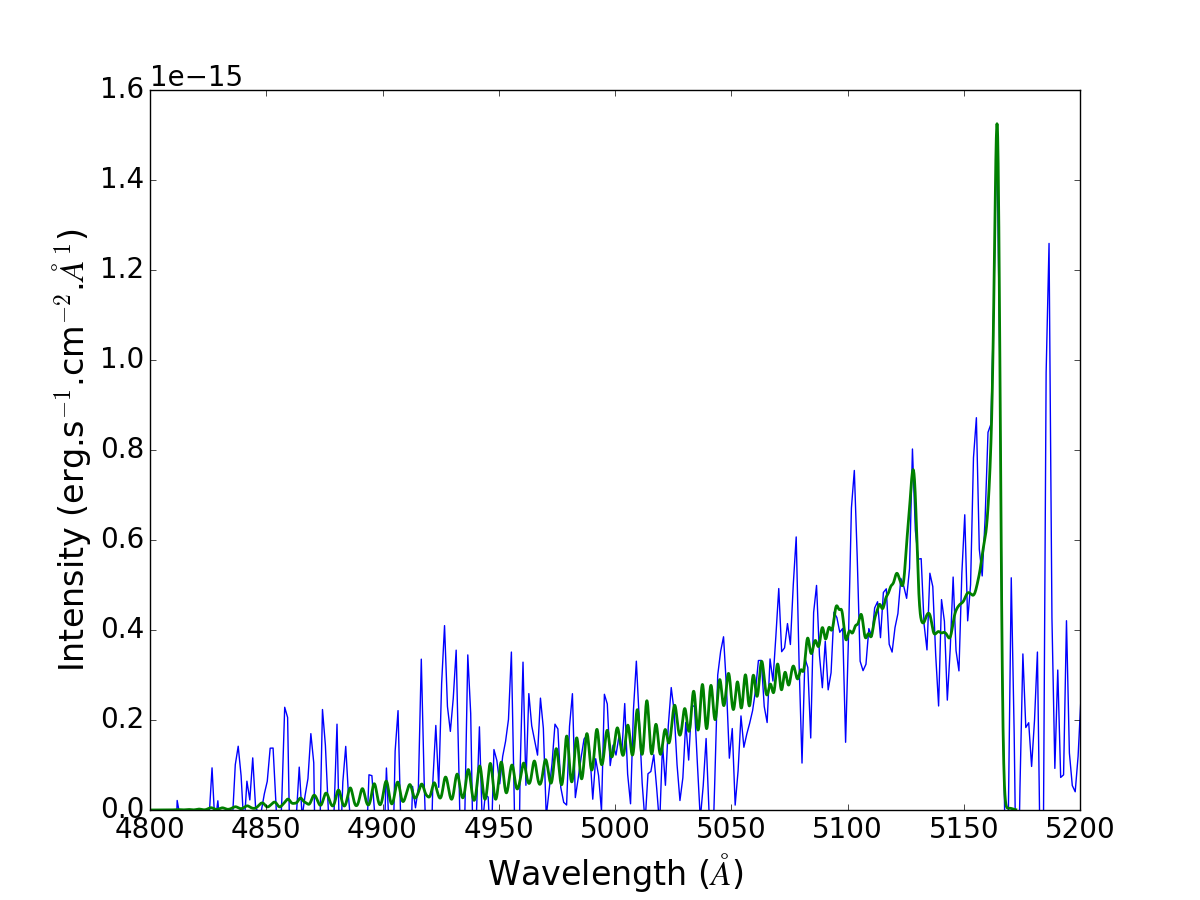}
    \includegraphics[width=\columnwidth]{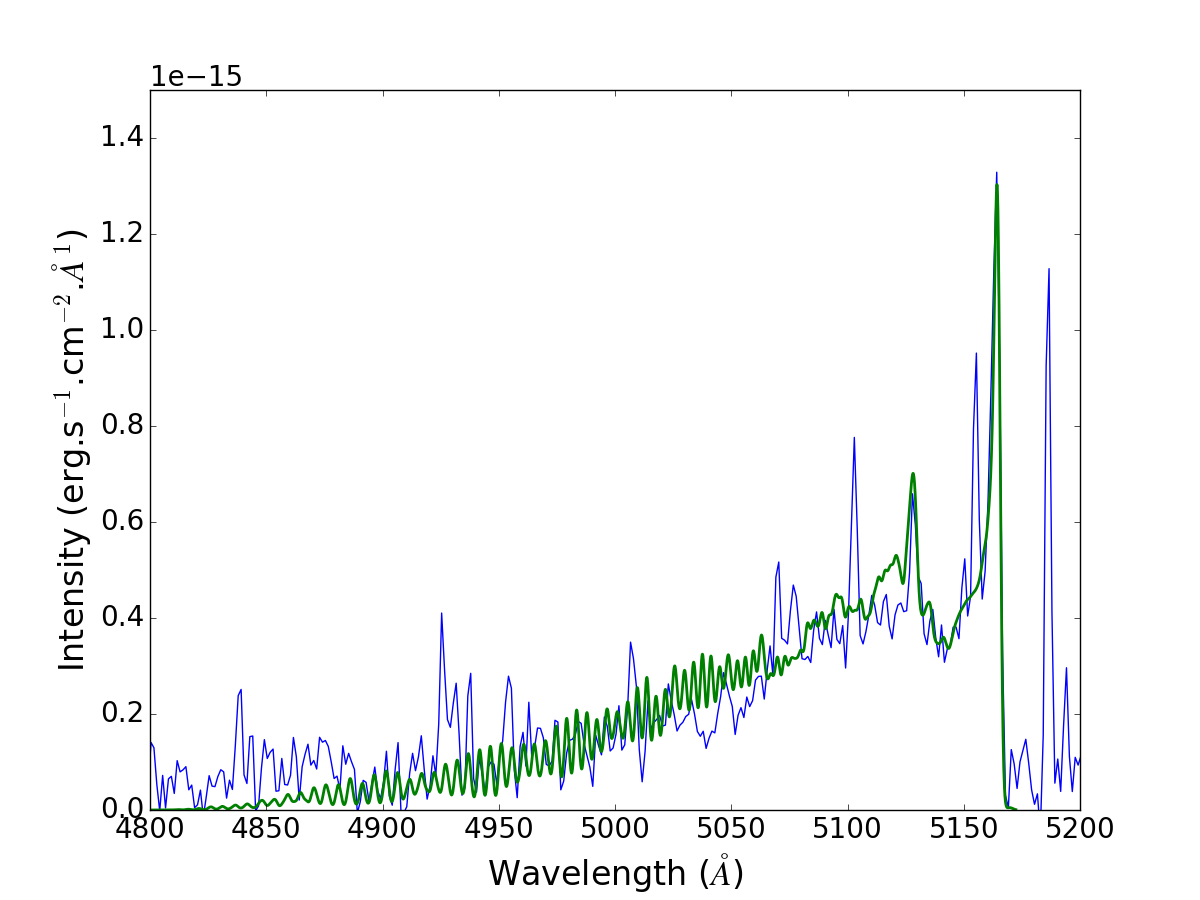}
    \includegraphics[width=\columnwidth]{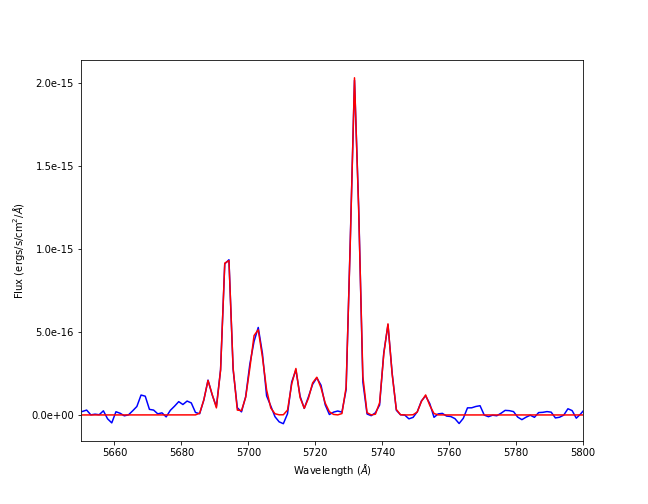}
    \caption{Detections in the MUSE spectra in Fig.~\ref{fig:spectrum}: \ce{C2} on 14-15 November (top), \ce{C2} on 26 November (middle), and \ce{NH2} on 26 November (lower). For \ce{C2}, observational data are shown in blue, the green spectra are fits computed with a fluorescence equilibrium. The observed \ce{NH2} (0,10,0) spectrum is shown in blue, with the gaussian fits to the component emission lines in red. Note that this is the \ce{NH2} (0,3,0) band in modern notation as explained by \citep{Cochran:2002}}
    \label{fig:C2_detection}
\end{figure}

\textbf{C$_2$ Swan ($\Delta v=0$) Band-Sequence}: In Fig.~\ref{fig:C2_detection}, we show the two detections of \ce{C2} in the MUSE spectra.
In order to interpret the results, we produced a model spectra assuming fluorescence equilibrium \citep{rousselot:2000}. 
Because the probability of intercombination transitions is low, this is a valid assumption, even if the timescale for \ce{C2} radicals to equilibrate is long. This model provided a better fit to the measured spectrum than a simple mix of two Boltzmann populations. 
This is most likely due to the large 10~arcsec radius of the field of view used to extract these spectra, which would provide ample time for the majority of radicals to reach equilibrium.  
From this synthetic spectrum, we computed the whole flux received from the complete Swan $\Delta v=0$ band, which are given in Table~\ref{tab:rates}.
The resulting production rate of $Q$(\ce{C2})$ = 1.1\times10^{24}$ mol s$^{-1}$ is above the upper limits from other facilities published previously; we discuss this further in \S~\ref{sec:discussion}. 
There are other notable lines and features in the spectrum. For instance, at 5188~\AA\, there is a feature due to the \ce{NH2} radical. 
At 1~au, this line is usually much weaker than the \ce{C2} feature.

\textbf{\ce{NH2} emission bands}:
Many bright emission lines of \ce{NH2} are present in the MUSE spectra, including the A–X (0,10,0)–(0,0,0) and A–X (0,8,0)–(0,0,0).
We fit the flux in the \ce{NH2}(0,10,0) band at 5680--5750~\AA\ by fitting individual emission features within the band with gaussians and summing (Fig.~\ref{fig:C2_detection}), as the \ce{NH2}(0-8-0) 6324--6337~\AA\ emission appeared too heavily contaminated by residuals from the sky subtraction of terrestrial \ce{[OI]} and/or cometary [OI] emission.
This is the first report of 2I's \ce{NH2} production rate.

\textbf{CN(1-0) red emission band}: We integrate the continuum-removed flux over 9100--9270~\AA\ with IRAF as for the \ce{NH2} band. This value was then corrected by a factor of 1.4 to account for the flux at longer wavelengths \citep{Fink:2009}, which were avoided because of the telluric noise, and potential second-order contamination.
The measurement uncertainties of $\sim10\%$ are more substantial than any uncertainty in the fluorescence efficiency ($g$-factor) known for the CN red band; \citet{Paganini:2016} find a $g$-factor of $41.9\times10^{-3}$ photons/molecule/s at 1 au, which is very close to the \citet{Fink:1994} value of $41\times10^{-3}$ photons/molecule/s. 

\textbf{CN(0-0) violet emission band}: As the red emission band of CN is not as frequently used to measure CN production rates, we provide  simultaneous or near-simultaneous constraints available from the TRAPPIST narrow-band imaging at 3890~\AA.
As shown in Table~\ref{tab:rates}, the CN production rates are consistent between these independent measurements from the red and violet bands. 
This verifies that it is reasonable to compare the production rates of gas species in 2I measured from the red CN band. 
We note that \citet[e.g.][]{Fink:2009} also measured Solar System comets in the red band of CN, and had production rates consistent with those of other surveys such as \citet{Ahearn:1995} and \citet{Cochran:2012}, so there is precedent for using the red CN band to examine cometary composition.

\section{Discussion}
\label{sec:discussion}

\begin{figure*}
    \includegraphics[width=\textwidth]{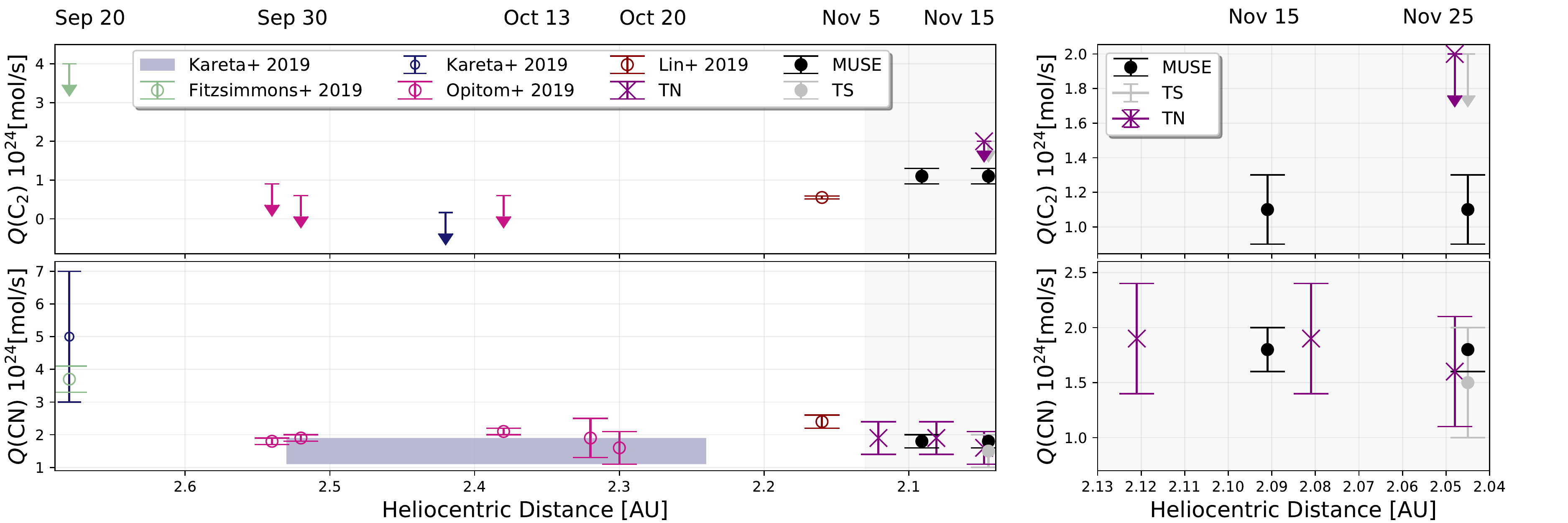}
    \caption{Pre-perihelion production rates of \ce{C2} (upper) and CN (lower) in 2I/Borisov in 2019 September--November; its perihelion is 2019 December 8. The left two panels show the measurements and upper limits in \citet{Fitzsimmons:2019}, \citet{Kareta:2019}, \citet{Opitom:2019-borisov}, \citet{Lin:2019}, and this work. The dark blue shaded region in the bottom left panel shows the continuous upper and lower limit measurements provided in \citet{Kareta:2019} for the production of CN during this period. The right two panels show our observations with MUSE, Trappist North and Trappist South.  }
    \label{fig:production_rates}
\end{figure*}

\begin{figure*}
    \includegraphics[width=\textwidth]{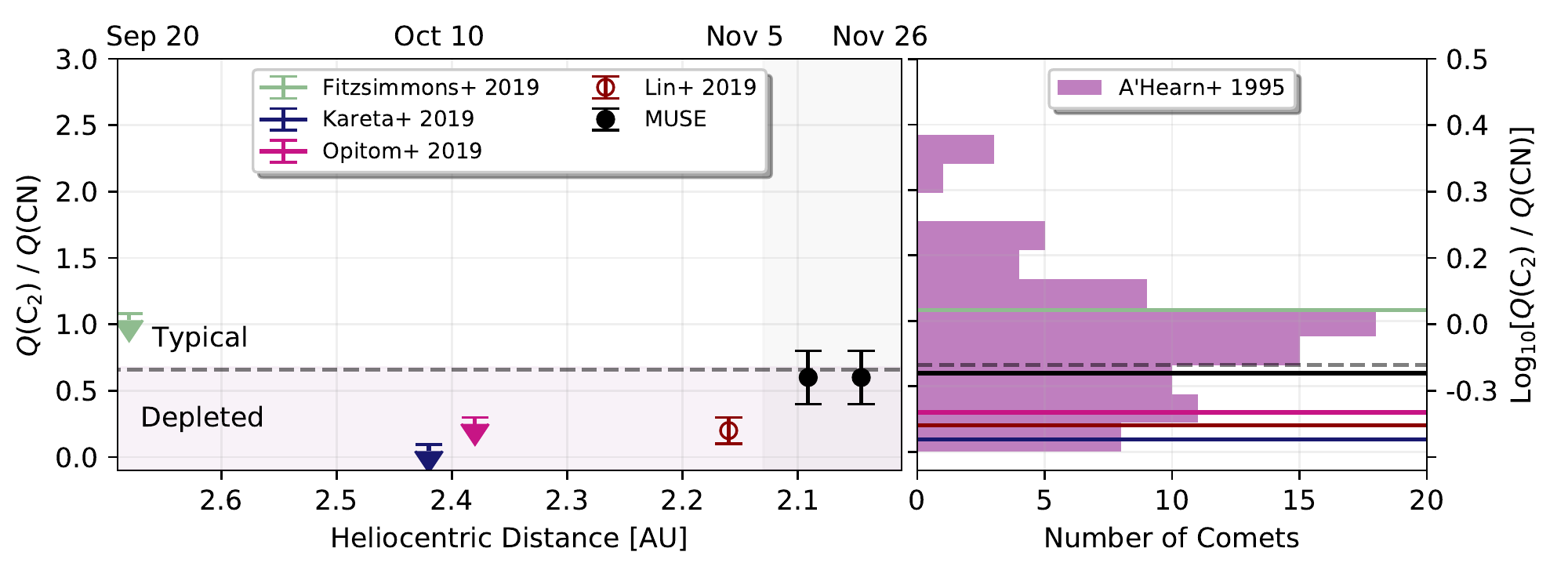}
    \caption{  Pre-perihelion ratio of the production rates of \ce{C2} and CN in 2I/Borisov in 2019 September--November; its perihelion is 2019 December 8. We show the measurements and upper limits in \citet{Fitzsimmons:2019}, \citet{Kareta:2019}, \citet{Opitom:2019-borisov}, \citet{Lin:2019}, and this work as a function of 2I/Borisov's heliocentric distance in the left panel. The two black points show the ratios measured with the  observations with MUSE on November 14-15 and November 25, when both CN and \ce{C2} are detected at high SNR. In the right panel, the purple histograms show the ratio of production rates for comets in the sample presented in \citet{Ahearn:1995}. The solid lines indicate the measured values and upper limits for Borisov, with the same color scheme as in the left panel. The black dashed line indicates the commonly accepted definition for carbon depleted comets, log[Q(\ce{C2})/Q(CN)]$\le -0.18$ }
    \label{fig:C2CN_ratio}
\end{figure*}

\begin{figure}
    \includegraphics[width=\columnwidth]{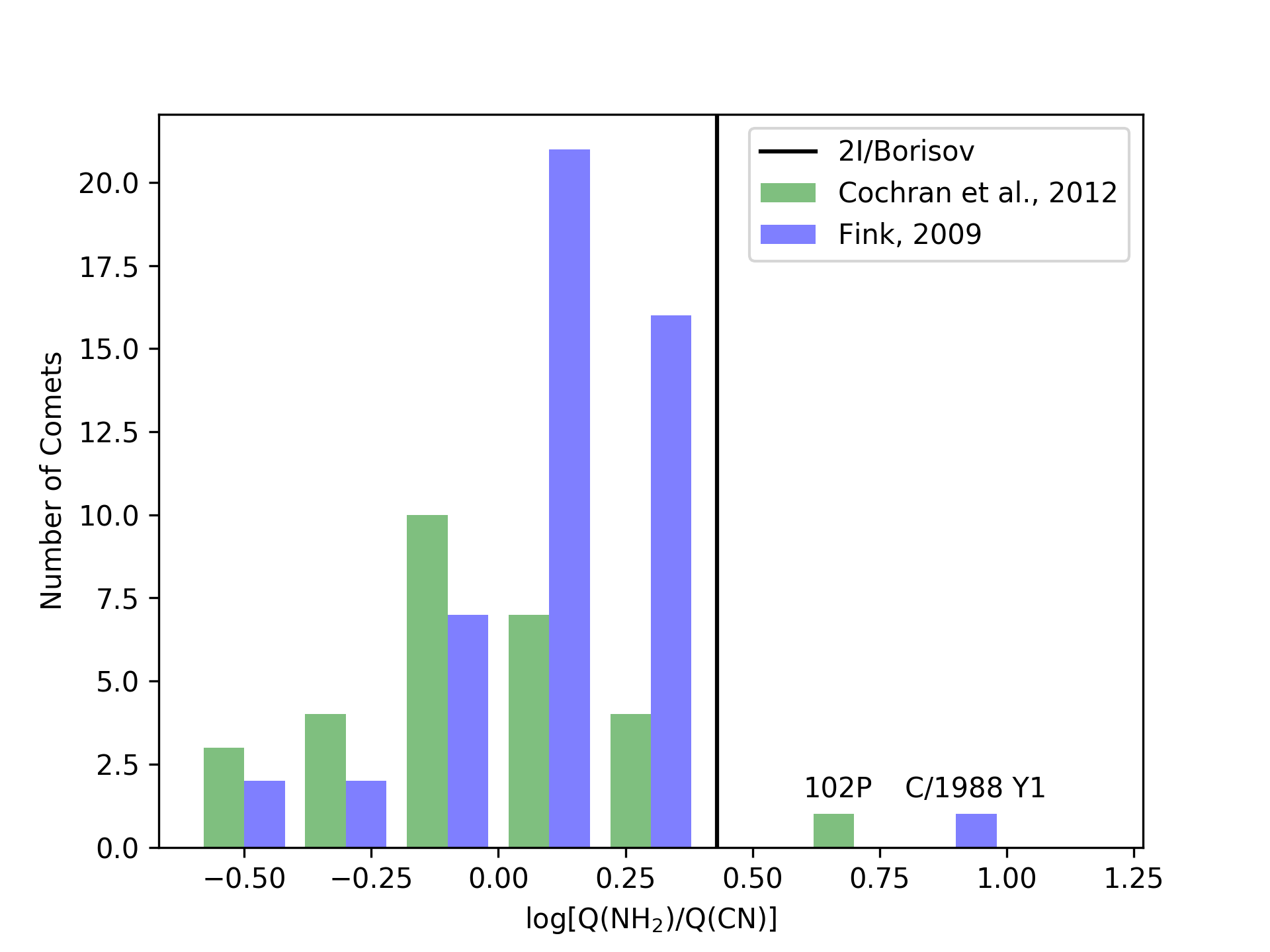}
    \caption{Ratio of the production rates of \ce{NH2} and CN for the comets with \ce{NH2} detections reported by \citet{Fink:2009, Cochran:2012}. The bin size is larger than the measurement uncertainties, where they exist. Only two comets (labelled) have a higher ratio than 2I/Borisov.
}
    \label{fig:NH2_ratio}
\end{figure}

2I's \ce{C2} production has started in earnest (Fig.~\ref{fig:production_rates}, upper left panel), given our high signal-to-noise prominent detection of \ce{C2} (Fig.~\ref{fig:C2_detection}).
In contrast, Fig.~\ref{fig:production_rates} (lower left panel) shows the CN production rate of 2I was stable within uncertainties between the end of September and the end of November. 
The higher CN production rate measurement on September 20 was under challenging conditions, since the sky background at this time was high; the SNR of \citet{Fitzsimmons:2019} is much better than the marginal detection of \citet{Kareta:2019}.
Records of the CN production rate at distances of several au are still sparse for dynamically new comets, though some have shown slow-increasing behaviour on inward approach \citep{Whipple:1978}.

The production of \ce{C2} relative to that of CN has increased relatively quickly in the coma of 2I. 
Fig.~\ref{fig:C2CN_ratio} (left panel) shows that the transition can be placed in the second week of November.
It may be that a new source of \ce{C2}, such as sublimation of organic-rich grains or another parent molecule, turned on once the comet came closer to the Sun. 
\citet{Cochran:2012} found that \ce{C2}/CN production rate ratios are constant with heliocentric distance (see their Fig. 10). 
In contrast, \citet{Langland-Shula:2011} concluded from their much smaller population snapshot that the \ce{C2}/CN in the coma of comets tends to increase as the heliocentric distance diminishes, with a best-fit power-law index of $2.2\pm 0.6$. 
However, most of their measurements were within 2 au.
Thus, the fast increase of 2I's production rate ratio may be notable: 2I was depleted in \ce{C2} on October 13, in comparison to comets observed at the same heliocentric distance \citep{Opitom:2019-borisov}. A suitable comparison may be C/2009 P1 Garradd. 
This comet had a sudden increase in \ce{C2} between 2.47 and 2.00 au inbound \citep{McKay:2013}, together with a huge change in CO/\ce{H2O} during its apparition \citep{Feaga:2014}.
 Unlike 2I \citep{Fitzsimmons:2019}, however,  C/2009 P1 was unusual in that it was an exceptionally dusty comet.
One interpretation of the change in composition of the
coma 
is that the subsurface layers of 2I that are now contributing to its activity are less carbon-depleted than 2I's uppermost layer. 
In this scenario, inbound sublimation removed the \ce{C2}-depleted layer. This would imply  that our November observations of the coma displays a more representative volatile composition.
It is also possible that a \ce{C2}-rich region has been recently exposed, implying a heterogeneity of the nucleus.

There have been few rates of change of the \ce{C2}/CN production rates measured for individual Solar System comets outside 2 au.  \citet{Langland-Shula:2011}'s four Q(\ce{C2})/Q(CN) measurements beyond 2 au are from only two different comets\footnote{017P and C/1995 O1 (Hale-Bopp)}. 
Intriguingly, \citet{Schulz:1998} reported a strong evolution in Q(\ce{C2})/Q(CN) for comet 46P/Wirtanen, with the comet being \ce{C2} depleted at 2 au, but having a normal abundance ratio at 1 au. However, \cite{Fink:2004} found this may be explained by varying Haser model parameters.
Thus, it is hard to robustly say if 2I's inbound abundance ratio change with $r_\mathrm{H}$ is unusually rapid. 
The lack of suitable Solar System comparison data reflects the challenge of measuring \ce{C2} at heliocentric distances greater than a few au. 
Additionally, the parameters (cf. Table~\ref{tab:model_parms}) and techniques used for computing production rates vary in the literature, and critically, these techniques are only resilient up to about 2 au.
Acquiring and analysing comet gas production rates in a uniform manner, particularly in detail when comets are inbound between 3 and 2 au, will be a fruitful goal for 8m+ telescopes in the future --- particularly given the expected increase in the rate of interstellar object discovery with the advent of the Rubin Observatory's\footnote{Formerly known as the Large Synoptic Survey Telescope.} surveying \citep{MoroMartin:2009,Cook:2016,Trilling:2017}.
We encourage the community to try to measure 2I's gas production ratios as often as possible through and beyond 3 au, to see if the behaviour is symmetric.

Comparing the ratio of \ce{C2} to CN production in 2I's coma is useful to define its relationship to Solar System comet populations.
Using a similar model and model parameters as we use in this work, \cite{Ahearn:1995} define a carbon-chain depleted comet as one with log[Q(\ce{C2})/Q(CN)] $<-0.18$.
Based on their non-detections or weak detections of \ce{C2}, previous studies concluded that 2I is depleted in carbon-chain species (Fig~\ref{fig:C2CN_ratio}, right panel), similar to some Solar System comets \citep{Kareta:2019, Opitom:2019-borisov, Lin:2019}.
According to \cite{Ahearn:1995}'s definition of carbon-chain depletion occurring at log[Q(\ce{C2})/Q(CN)] $\leq-0.21$, 2I is still carbon-chain depleted at the epochs we measure, but not nearly as much as was first thought (Fig.~\ref{fig:C2CN_ratio}, right panel).
In our Solar System, both Oort-family and Jupiter-family comets show carbon-chain depletion consistent with 2I's depletion. Therefore, 2I is not directly analogous to  any one  dynamical family of comets. Unfortunately, it will not be possible to identify a likely region of origin in its star's protoplanetary disk via comparison with the formation of Solar System comets. 

The detection of \ce{NH2} in the MUSE data is significant, because \ce{NH2} is a dissociation product of \ce{NH3}, which is a primary nitrogen reservoir in solar system comets.
\ce{NH2} is prominent in our spectra, yielding a ratio for 2I of Q(\ce{NH2})/Q(CN)=2.7. Only a handful of optical studies of Solar System comet composition include measurements of the \ce{NH2} production rates: \citet{Beaver:1990} (6 comets at 0.65-1.8~au), \citet{Fink:2009} (50 comets), and \citet{Cochran:2012} (26 comets with \ce{NH2} detections that were observed on at least three nights, although with some overlap with the sample of \citet{Fink:2009}). 
Fig.~\ref{fig:NH2_ratio} shows that the \ce{NH2} to CN production ratio that we measure for 2I at 2.02 au of log[Q(\ce{NH2})/Q(CN)]$=0.43$ is higher than almost all of the comets in the \cite{Cochran:2012} and \cite{Fink:2009} data sets, with the exception only of comets 102P and C/1988 Y1.
\citet{Cochran:2012} report a mean log production rate ratio with respect to CN of $0.09 \pm 0.25$ for the 26 comets with \ce{NH2} in their restricted dataset; thus 2I is within $1.4\sigma$.
While there is no notable trend of the \ce{NH2}/CN production ratio currently known between 1 and 2 au, we note that for \citet{Fink:2009}'s data set, only two comets have \ce{NH2} production rates measured beyond 2 au (one is relatively rich and one is poor in \ce{NH2}), and for \citet{Cochran:2012}, there are no values of the \ce{NH2}/CN ratio beyond 2 au.
The current statistics are insufficient to refine the comparison of 2I's \ce{NH2}/CN production ratio more finely relative to a dynamical origin as Oort or Jupiter family comets. Comparison to literature values requires care, because the MUSE spectrum has been extracted close to the nucleus (within 10 arcsec), and \ce{NH2} has a much shorter lifetime than \ce{C2} and CN, while usually long-slit spectra are extracted on a larger area ($\sim1$ arcmin).
Nuclei size should also be considered, as at 0.2--0.5 km, 2I is smaller than most well-studied comets to date.
Nevertheless, 2I is relatively rich in \ce{NH2} compared to the available bulk of Solar System comets.

The kinship of 2I's physical properties implies an origin system that had a planetesimal-forming disk with many chemical similarities to our own.
Many difficulties remain in linking the span of processes from disk to comets.
Our own system's cometary populations appear largely homogeneous, with subtle compositional variation \citep[c.f.][]{Cochran:2012}.
Identifying nuanced differences that could produce variation like the \ce{NH2} enrichment we observe in 2I will be an ongoing task.

\section{Conclusion}
\label{sec:conclusion}
 
We report the detection of \ce{C2}, \ce{NH2} and the red bandhead of CN in 2I/Borisov, the first interstellar object detected passing through the Solar System with prominent cometary activity. 
We present measurements from the MUSE ESO/VLT integral-field spectrograph taken on three nights in November 2019. 
These measurements are the first of a larger observing program that is monitoring 2I through its perihelion passage. 
We combine these data with complementary observations from the two TRAPPIST telescopes in order to constrain the production rates of various species in the coma of 2I.
The CN production rates measured near-simultaneously from the violet and red bandheads of CN emission are consistent within uncertainties.
As of 2019 November 26, the dust continuum of 2I has a spectral slope in the optical of $11.0\pm 0.2$ \%/$10^3$ \AA, consistent with earlier reports.
The \ce{C2} production of 2I has activated, and a changing \ce{C2}/CN production ratio is apparent.
At this epoch, 2I is only slightly depleted in \ce{C2} at Q(\ce{C2})/Q(CN)=0.61, but it is  rich in \ce{NH2} relative to almost all Solar System comets, at Q(\ce{NH2})/Q(CN)=2.7.
The known sample of Solar System comets observed at 2 au and beyond is sufficiently small to hamper comparison, particularly in the production rates for \ce{NH2} relative to that of CN in carbon-chain depleted comets.
2I's chemical composition retains many similarities to the broader population of Solar System comets.

\acknowledgements

M.T.B. and A.F. appreciate support from UK STFC grant ST/P0003094/1. 
M.M.K. was supported by NASA Solar System Observations grant 80NSSC18K0856. D.S. is supported by the National Aeronautics and Space Administration through the NASA Astrobiology Institute under Cooperative Agreement Notice NNH13ZDA017C issued through the Science Mission Directorate.
M.M. was supported by the National Aeronautics and Space Administration under Grant No. 80NSSC18K0849 issued through the Planetary Astronomy Program. 
A.G.L. was supported by the European Research Council (ERC) under grant agreement No. 802699.

Based on observations collected at the European Southern Observatory under ESO programme 2103.C-5070. 
We thank the ESO staff, particularly Henri Boffin, Bin Yang, Diego Parraguez, Edmund Christian Herenz, Fuyan Bian, and Israel Blanchard, for their help in the acquisition of these observations.

TRAPPIST is a project funded by the Belgian Fonds (National) de la Recherche Scientifique (F.R.S.-FNRS) under grant FRFC 2.5.594.09.F and the ARC grant for Concerted Research Actions, financed by the Wallonia-Brussels Federation. TRAPPIST-North is a project funded by the University of Liege, in collaboration with Cadi Ayyad University of Marrakech (Morocco). E.J is a F.R.S.-FNRS Senior Research Associate.

\bibliography{paper}{}
\bibliographystyle{aasjournal}

\end{document}